\documentclass[journal]{vgtc}                

\usepackage{xspace}
\usepackage[dvipsnames]{xcolor}
\usepackage{enumitem}
\usepackage{wrapfig}
\usepackage[normalem]{ulem}

\usepackage{amssymb}
\usepackage{amsmath}
\usepackage{mathtools}

\usepackage{dblfloatfix}    

\usepackage{array}
\newcolumntype{L}[1]{>{\raggedright\let\newline\\\arraybackslash\hspace{0pt}}m{#1}}
\newcolumntype{C}[1]{>{\centering\let\newline\\\arraybackslash\hspace{0pt}}m{#1}}
\newcolumntype{R}[1]{>{\raggedleft\let\newline\\\arraybackslash\hspace{0pt}}m{#1}}
\newcolumntype{P}[1]{>{\raggedright}p{#1}}

\newcommand{\rem}[1]{}

\definecolor{urlBlue}{HTML}{2196F3}
\newcommand{\urlColor}[1]{{\color{urlBlue}{#1}\normalfont}}

\newcommand{\model}{\textsc{InceptionV1}\xspace}
\newcommand{\system}{\textsc{Summit}\xspace}
\newcommand{\embedding}{Embedding View\xspace}
\newcommand{\sidebar}{Class Sidebar\xspace}
\newcommand{\graph}{Attribution Graph View\xspace}
\newcommand{\acta}{activation aggregation\xspace}
\newcommand{\infa}{neuron-influence aggregation\xspace}

\newcommand{\ag}{attribution graph\xspace}
\newcommand{\ags}{attribution graphs\xspace}

\newcommand{\AGs}{Attribution Graphs\xspace}
\newcommand{\Ag}{Attribution graph\xspace}
\newcommand{\Ags}{Attribution graphs\xspace}
\newcommand{\A}{$A$\xspace}

\newcommand{\demoURL}{\url{https://fredhohman.com/summit/}\xspace}

\definecolor{main}{HTML}{1A237E}
\definecolor{mainLight}{HTML}{7986CB}

\definecolor{techGreen}{HTML}{388E3C}
\definecolor{techPurple}{HTML}{1A237E}
\definecolor{techBlue}{HTML}{42A5F5}
\definecolor{techRed}{HTML}{EF5350}

\definecolor{mixedBlue}{HTML}{1EB1FC}
\definecolor{mixedGreen}{HTML}{69B96D}

\definecolor{unexOrange}{HTML}{FFB300}

\definecolor{bearBrown}{HTML}{A97138}
\definecolor{bearBlack}{HTML}{444444}
\definecolor{bearShare}{HTML}{999999}

\newcommand{\blackBear}{{\textit{\textbf{\textcolor{bearBlack}{black bear}}}}\xspace}
\newcommand{\brownBear}{{\textit{\textbf{\textcolor{bearBrown}{brown bear}}}}\xspace}
\newcommand{\bearness}{{\textit{\textbf{\textcolor{bearShare}{bear-ness}}}}\xspace}

\newcommand{\actStep}[1]{{\textbf{(\textcolor{techGreen}{#1})}\normalfont}}
\newcommand{\infStep}[1]{{\textbf{(\textcolor{techPurple}{#1})}\normalfont}}

\usepackage{tcolorbox}
\definecolor{tagbordercolor}{HTML}{B0BEC5}
\definecolor{tagbgcolor}{HTML}{ECEFF1}

\newtcbox{\captag}{nobeforeafter, colframe=tagbordercolor,
colback=tagbgcolor, boxrule=0.5pt, arc=1pt,
 boxsep=0pt,left=2pt,right=2pt,top=1.5pt,bottom=2pt,tcbox raise base}

\ifpdf
  \pdfoutput=1\relax                   
  \usepackage{graphicx}                
  \DeclareGraphicsExtensions{.pdf,.png,.jpg,.jpeg} 
\else
  \usepackage{graphicx}                
  \DeclareGraphicsExtensions{.eps}     
\fi%

\graphicspath{{figures/}{pictures/}{images/}{./}} 

\usepackage{microtype}                 
\PassOptionsToPackage{warn}{textcomp}  
\usepackage{textcomp}                  
\usepackage{mathptmx}                  
\usepackage{times}                     
\usepackage{cite}                      
\usepackage{tabu}                      
\usepackage{booktabs}                  
\usepackage{enumitem}

\onlineid{0}

\vgtccategory{Research}
\vgtcpapertype{algorithm/technique}

\title{\system: Scaling Deep Learning Interpretability by\\Visualizing Activation and Attribution Summarizations}

\author{Fred Hohman, Haekyu Park, Caleb Robinson, and Duen Horng (Polo) Chau}
\authorfooter{
\item
 Fred Hohman, Haekyu Park, Caleb Robinson, and Duen Horng Chau are with Georgia Tech. E-mail: \{fredhohman, haekyu, dcrobins, polo\}@gatech.edu.
}

\shortauthortitle{Hohman \MakeLowercase{\textit{et al.}}: \system}

\abstract{Deep learning is increasingly used in decision-making tasks.
However, understanding how neural networks produce final predictions remains a fundamental challenge.
Existing work on interpreting neural network predictions for images often focuses on explaining predictions for single images or neurons.
As predictions are often computed from millions of weights that are optimized over millions of images, such explanations can easily miss a bigger picture.
We present \system, an interactive system that scalably and systematically summarizes and visualizes what features a deep learning model has learned and how those features interact to make predictions.
\system introduces two new scalable summarization techniques: 
(1) \textit{\acta} discovers important neurons, and 
(2) \textit{\infa} identifies relationships among such neurons. 
\system combines these techniques to create the novel \textit{attribution graph} that reveals and summarizes crucial neuron associations and substructures that contribute to a model's outcomes.
\system scales to large data, such as the ImageNet dataset with 1.2M images, and leverages neural network feature visualization and dataset examples to help users distill large, complex neural network models into compact, interactive visualizations.
We present neural network exploration scenarios where \system helps us discover multiple surprising insights into a prevalent, large-scale image classifier's learned representations and informs future neural network architecture design.
The \system visualization runs in modern web browsers and is open-sourced.
}


\CCScatlist{ 
 \CCScat{K.6.1}{Management of Computing and Information Systems}%
{Project and People Management}{Life Cycle};
 \CCScat{K.7.m}{The Computing Profession}{Miscellaneous}{Ethics}
}

\teaser{
  \centering
  \includegraphics[width=\linewidth]{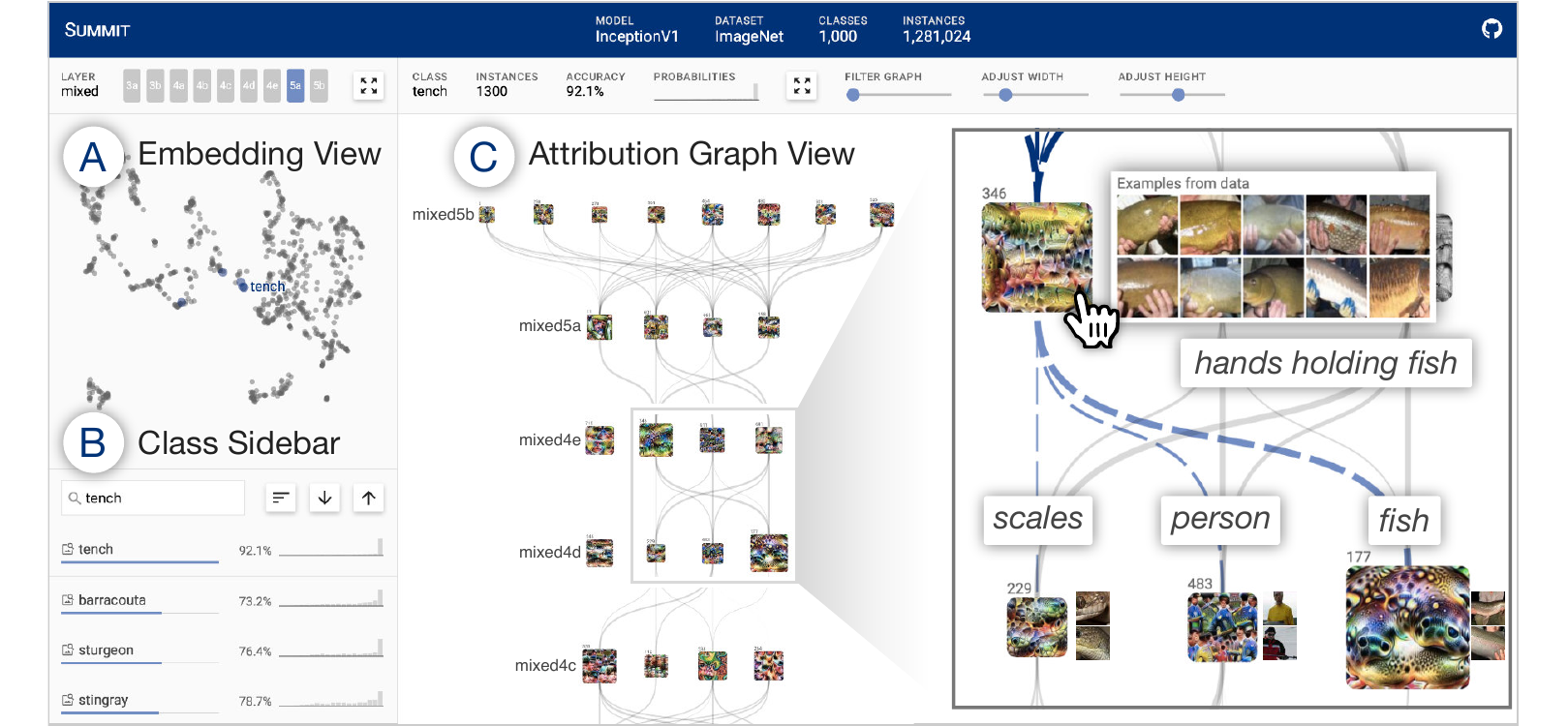}
   \caption{
        With Summit, users can scalably summarize and interactively interpret deep neural networks
        by visualizing \textit{what} features a network detects and \textit{how} they are related.
        In this example, \model accurately classifies images of \textit{\textbf{tench}} (yellow-brown fish).
        However, \system reveals surprising associations in the network (e.g., using parts of people) that contribute to its final outcome: the ``tench'' prediction is dependent on an intermediate ``hands holding fish'' feature (right callout), which is influenced by lower-level features like \textit{``scales,''} \textit{``person,''} and \textit{``fish''}.
        \textbf{(\textcolor{main}{A}) \embedding} summarizes all classes' aggregated activations using dimensionality reduction.
        \textbf{(\textcolor{main}{B}) \sidebar} enables users to search, sort, and compare all classes within a model.
        \textbf{(\textcolor{main}{C}) \graph} visualizes highly activated neurons as vertices (``scales,'' ``fish'') and their most influential connections as edges (dashed purple edges).
 }
	\label{fig:teaser}
}

\vgtcinsertpkg

\begin{document}


\firstsection{Introduction}

\maketitle


Deep learning is increasingly used in decision-making tasks, due to its high performance on previously-thought hard problems and a low barrier to entry for building, training, and deploying neural networks.
Inducing a model to discover important features from a dataset is a powerful paradigm, yet this introduces a challenging \textit{interpretability} problem---it is hard for people to understand what a model has learned.
This is exacerbated in situations where a model could have impact on a person's safety, financial, or legal status~\cite{council2016gdpr}. 
Definitions of interpretability center around \textit{human understanding}, but they vary in the aspect of the model to be understood: its internals~\cite{gilpin2018explaining}, operations~\cite{biran2017explanation}, mapping of data~\cite{montavon2017methods}, or representation~\cite{ribeiro2016should}.
Although recent work has begun to operationalize interpretability~\cite{hohman2019gamut}, a formal, agreed-upon definition remains open~\cite{doshi2017towards, lipton2016mythos}.

Existing work on interpreting neural network predictions for images often focuses on explaining predictions for single images or neurons~\cite{selvaraju2017grad, smilkov2017smoothgrad, olah2017feature, olah2018building}.
As large-scale model predictions are often computed from millions of weights optimized over millions of images, such explanations can easily miss a bigger picture.
Knowing how entire classes are represented inside of a model is important for trusting a model's predictions and deciphering what a model has learned~\cite{ribeiro2016should}, since these representations are used in diverse tasks like detecting breast cancer~\cite{liu2018artificial, steiner2018impact}, predicting poverty from satellite imagery~\cite{jean2016combining}, defending against adversarial attacks~\cite{das2018shield}, transfer learning~\cite{zamir2018taskonomy, pan2010survey}, and image style transfer~\cite{gatys2016image}.
For example, high-performance models can learn unexpected features and associations that may puzzle model developers.
Conversely, when models perform poorly, developers need to understand their causes to fix them \cite{kahng2018activis,ribeiro2016should}.
As demonstrated in \autoref{fig:teaser}, \model, a prevalent, large-scale image classifier, accurately classifies images of \textit{\textbf{tench}} (yellow-brown fish).
However, our system, \system, reveals surprising associations in the network that contribute to its final outcome:
\textit{\textbf{tench}} is dependent on an intermediate person-related ``hands holding fish'' feature (right callout) influenced by lower-level features like \textit{``scales,''} \textit{``person,''} and \textit{``fish''}.
There is a lack of research in developing scalable summarization and interactive interpretation tools that simultaneously reveal important neurons and their relationships.
\system aims to fill this critical research gap.

\begin{figure}[t]
 \centering
 \includegraphics[width=0.85\linewidth]{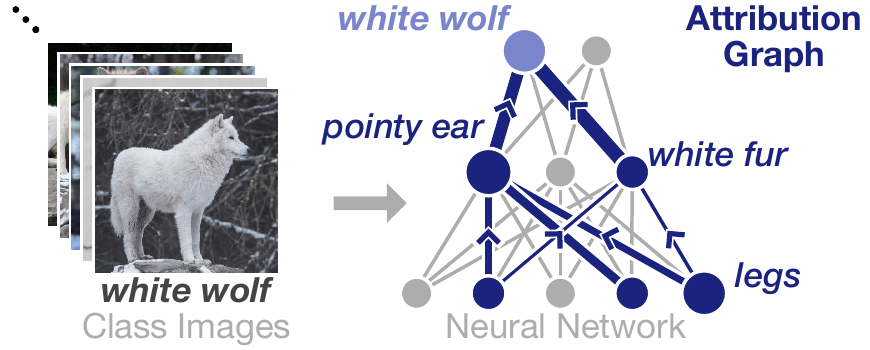}
 \vspace{-3mm}
 \caption{A high-level illustration of how we take thousands of images for a given class, e.g., images from \textbf{\textit{white wolf}} class, compute their top activations and attributions, and combine them to form an \textcolor{main}{\textbf{attribution graph}} that shows how lower-level features (``legs'') contribute to higher-level ones (``white fur''), and ultimately the final outcome.}
 \label{fig:attribution-graph}
 \vspace{-5mm}
\end{figure}

\smallskip
\noindent \textbf{Contributions.} In this work, we contribute:

\begin{itemize}[topsep=1mm, itemsep=0mm, parsep=1mm, leftmargin=3mm]

\item 
\textbf{\system, an interactive system for scalable summarization and interpretation} 
for exploring entire learned classes in prevalent, large-scale image classifier models, such as \model~\cite{szegedy2015going}.
\system leverages neural network feature visualization~\cite{olah2017feature, nguyen2017plug, mordvintsev2015inceptionism, simonyan2013deep, erhan2009visualizing}
and dataset examples to distill large, complex neural network models into compact, interactive graph visualizations (\autoref{sec:ui}).

\item
\textbf{Two new scalable summarization techniques}
for deep learning interpretability:
(1) \textit{\acta} discovers important neurons (\autoref{subsec:activation}), and 
(2) \textit{\infa} identifies relationships among such neurons (\autoref{subsec:influence}). 
These techniques scale to large data, e.g., ImageNet ILSVRC 2012 with 1.2M images~\cite{russakovsky2015imagenet}.

\item
\textbf{\Ag, a novel way to summarize and visualize entire classes},
by combining our two scalable summarization techniques to reveal crucial neuron associations and substructures that contribute to a model's outcomes, simultaneously highlighting \textit{what} features a model detects, and \textit{how} they are related (\autoref{fig:attribution-graph}).
By using a graph representation, we can leverage the abundant research in graph algorithms to extract \ags from a network that show neuron relationships and substructures within the entire neural network that contribute to a model's outcomes (\autoref{subsec:combine}).

\item 
\textbf{An open-source, web-based implementation}
that broadens people's access to interpretability research 
without the need for advanced computational resources.
Our work joins a growing body of open-access research that aims to use interactive visualization to explain complex inner workings of modern machine learning techniques~\cite{olah2017research, kahng2019gan, smilkov2017direct}.
Our computational techniques for aggregating activations, aggregating influences, generating \ags and their data, as well as the \system visualization, are open-sourced\footnote{
Visualization: \url{https://github.com/fredhohman/summit}.\\
Code: \url{https://github.com/fredhohman/summit-notebooks}.\\
Data: \url{https://github.com/fredhohman/summit-data}.
\\}.
The system is available at the following public demo link: \urlColor{\demoURL}.

\end{itemize}

\noindent\textbf{Neural network exploration scenarios.}
Using \system, we investigate how a widely-used computer vision model hierarchically builds its internal representation that has merely been illustrated in previous literature.
We present neural network exploration scenarios where \system helps us discover multiple surprising insights into a prevalent, large-scale image classifier's learned representations and informs future neural network architecture design (\autoref{sec:results}).

\smallskip
\noindent\textbf{Broader impact for visualization in AI.}
We believe our summarization approach that builds entire class representations is an important step for developing higher-level explanations for neural networks. 
We hope our work will inspire deeper engagement from both the information visualization and machine learning communities to further develop human-centered tools for artificial intelligence~\cite{olah2017research, abdul2018trends}.

\section{Background for Neural Network Interpretability}
\label{sec:background}

Typically, a neural network is given an input data instance (e.g., an image) and computes transformations on this instance until ultimately producing a probability for prediction.
Inside the network at each layer, each neuron (i.e., channel) detects a particular feature from the input.
However, since deep learning models learn these features through training, research in interpretability investigates how to make sense of what specific features a network has detected.
We provide an overview of existing activation-based methods for interpretability, a common approach to understand how neural networks operate internally that considers the magnitude of each detected feature inside hidden layers.

\subsection{Understanding Neuron Activations}
 
\textbf{Neuron activations as features for interpretable explanations.}
There have been many approaches that use neuron activations as features for interpretable explanations of neural network decisions. 
TCAV vectorizes activations in each layer and uses the vectors in a binary classification task for determining an interpretable concept's relevance (e.g., striped pattern) in model's decision for a specific class (e.g., zebra)~\cite{kim2017interpretability}.
Network Dissection~\cite{bau2017network} and Net2Vec~\cite{fong2018net2vec} propose methods to quantify interpretability by measuring alignment between filter activations and concepts.
ActiVis visualizes activation patterns in an interactive table view, where the columns are neurons in a network and rows are data instances~\cite{kahng2018activis}.
This table unifies instance-level and subset-level analysis, which enables users to explore inside neural networks and visually compare activation patterns across images, subsets, and classes.

\begin{figure}[b]
 \centering
 \includegraphics[width=0.94\columnwidth]{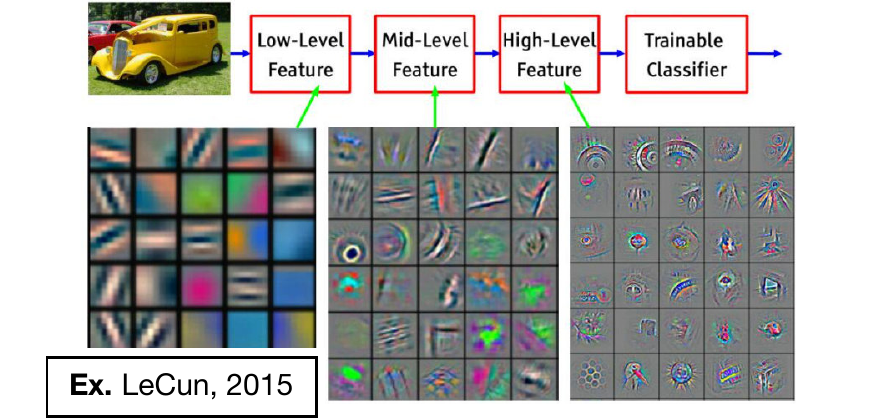}
 \vspace{-3mm}
 \caption{
    A common, widely shared example illustrating how neural networks learn hierarchical feature representations.
    Our work crystallizes these illustrations by systematically building a graph representation that describe \textit{what} features a model has learned and \textit{how} they are related.
    We visualize features learned at individual neurons and connect them to understand how high-level feature representations are formed from lower-level features.
    \textbf{Ex.} taken from Yann LeCun, 2015.
 }
 \label{fig:background}
\end{figure}

\textbf{Visualizing neurons with their activation.}
Instead of only considering the magnitude of activations, another technique called feature visualization algorithmically generates synthetic images that maximize a particular neuron~\cite{olah2017feature, nguyen2017plug, mordvintsev2015inceptionism, simonyan2013deep, erhan2009visualizing,carter2019activation}.
Since these feature visualizations optimize over a single neuron, users can begin to decipher what feature a single neuron may have learned.
These techniques have provided strong evidence of how neural networks build their internal hierarchical representations~\cite{zeiler2014visualizing}.
\autoref{fig:background} presents widely shared examples of how neural networks learn hierarchical features by showing neuron feature visualizations. 
It is commonly thought that neurons in lower layers in a network learn low-level features, such as edges and textures, while neurons in later layers learn more complicated parts and objects, like faces (\autoref{fig:background}).
In our work, we crystallize this belief by leveraging feature visualization to identify what features a model has detected, and how they are related.

\subsection{Towards Higher-level Deep Learning Interpretation}
It is not uncommon for modern, state-of-the-art neural networks to contain hundreds of thousands of neurons; visualizing all of them is ineffective. 
To address this problem, several works have proposed to extract only ``important'' neurons for a model's predictions~\cite{carter2019activation, liu2018analyzing, olah2018building}.
For example, Blocks, a visual analytics system, shows that class confusion patterns follow a hierarchical structure over the classes~\cite{bilal2018convolutional}, and Activation Atlases, large-scale dimensionality reductions, show many averaged activations~\cite{carter2019activation}.
Both visualizations reveal interesting properties of neural networks.
However, they either (1) consider activations independent of their learned connections, (2) depend on randomized sampling techniques, or (3) are computationally expensive.
\system addresses these issues by:
(1) combining both activations and relationships between network layers, as only knowing the most important neurons is not sufficient for determining how a model made a prediction---relationships between highly contributing neurons are key to understanding how learned features are combined inside a network; (2) leveraging entire datasets; and (3) integrating scalable techniques.

Since feature visualization has shown that neurons detect more complicated features towards a network's output, it is reasonable to hypothesize that feature construction is the collaborative combination of many different features from previous layers~\cite{selvaraju2018choose, bau2017network, fong2018net2vec}.
Our visualization community has started to investigate this hypothesis.
For example, one of the earlier visual analytics approaches, CNNVis, derives neuron connections for model diagnosis and refinement, but did not scale to large datasets with many classes~\cite{liu2017towards}.
In the context of adversarial machine learning, AEVis uses backpropagation to identify where in a network the data paths of a benign and attacked instance diverge~\cite{liu2018analyzing}.
AEVis demonstrates its approach on single and small sets of images; it is unclear how the approach's integral approximation optimization techniques scale to large, entire datasets, such as ImageNet.
Another example, Building Blocks, proposes to use matrix factorization to group sets of neurons together within a layer and derive ``compatible'' neuron groups across layers~\cite{olah2018building}; however, the work suggests uncertainty in the proposed formulation.
Our work draws inspirations from the above important prior research in neural network visualization.
Our method makes advances to scale to large million-image datasets, providing new ways to interpret entire classes (vs. single-image explanations) by aggregating activations and influences across the model.

\subsection{Visual Analytics for Interpretability}
To better facilitate interpretability, interactive visual analytics solutions have been proposed to help different user groups interpret models using a variety of interactive and visualization techniques~\cite{hohman2018visual}.
Predictive visual analytics supports experts conducting performance analysis of machine learning models by visualizing distributions of predicted instances, computing feature importance, and directly inspecting model and instance errors to support debugging~\cite{wongsuphasawat2018visualizing, lu2017state, hohman2019gamut, ren2017squares, amershi2015modeltracker}.
Interactive visualization for explaining models to non-experts using direct manipulation has also seen attention due to the pervasiveness of machine learning in modern society and general interest from the public~\cite{smilkov2017direct, kahng2019gan, harley2015isvc}.

\section{Design Challenges}
\label{sec:challenges}

Our goal is to build an interactive visualization tool for users to better understand how neural networks build their hierarchical representation.
To develop our summarization techniques and design \system, we identified five key challenges.

\begin{itemize}[label=C\arabic*.,itemsep=0mm, topsep=2mm]

\item[\textbf{C1.}]
\captag{\sffamily \textsc{\small{Scalability}}}
\textbf{Scaling up explanations and representations to entire classes, and ultimately, datasets of images.}
Much of the existing work on interpreting neural networks focuses on visualizing the top independent activations or attributions for a single image~\cite{selvaraju2017grad, smilkov2017smoothgrad, olah2017feature, olah2018building}.
While this can be useful, it quickly becomes tiresome to inspect these explanations for more than a handful of images.
Furthermore, since every image may contain different objects, to identify which concepts are representative of the learned model for a specific class, users must compare many image explanations together to manually find commonalities.

\item[\textbf{C2.}]
\captag{\sffamily \textsc{\small{Influence}}}
\textbf{Discovering influential connections in a network that most represents a learned class.}
In dense neural network models, scalar edge weights directly connect neurons in a previous layer to neurons in a following layer; in other words, the activation of single neuron is expressed as a weighted sum of the activations from neurons in the previous layer~\cite{goodfellow2016deep}.
However, this relationship is more complicated in convolutional neural networks.
Images are convolved to form many 2D activation maps, that are eventually summed together to form the next layers activations.
Therefore, it becomes non-trivial to determine the effect of a single convolutional filter's effect on later layers.

\item[\textbf{C3.}]
\captag{\sffamily \textsc{\small{Visualization}}}
\textbf{Synthesizing meaningful, interpretable visualizations with important channels and influential connections.}
Given a set of top activated neurons for a collection of images, and the impact convolutional filters have on later layers, how do we combine these approaches to form a holistic explanation that describes an entire class of images?
Knowing how entire classes are represented inside of a model is important for trusting a model's predictions~\cite{ribeiro2016should}, aiding decision making in disease diagnosis~\cite{liu2018artificial, steiner2018impact}, devising security protocols~\cite{das2018shield}, and fixing under-performing models \cite{kahng2018activis, ribeiro2016should}.

\item[\textbf{C4.}]
\captag{\sffamily \textsc{\small{Interaction}}}
\textbf{Interactive exploration of hundreds of learned class representations in a model.}
How do we support interactive exploration and interpretation of hundreds or even thousands of classes learned by a prevalent, large-scale deep learning model?
Can an interface support both high-level overviews of learned concepts in a network, while remaining flexible to support filtering and drilling down into specific features?
Whereas \textbf{C1} focuses on the summarization approaches to scale up representations, this challenge focuses on interaction approaches for users to work with the summarized representations.

\item[\textbf{C5.}]
\captag{\sffamily \textsc{\small{Research Access}}}
\textbf{High barrier of entry for understanding large-scale neural networks.}
Currently, deep learning models require extensive computational resources and time to train and deploy.
Can we make understanding neural networks more accessible without such resources, so that everyone has the opportunity to learn and interact with deep learning interpretability?

\end{itemize}

\section{Design Goals}
\label{sec:goals}

Based on the identified design challenges (\autoref{sec:challenges}), we distill the following main design goals for \system, an interactive visualization system for summarizing what features a neural network has learned.

\begin{itemize}[label=G\arabic*.,itemsep=0mm, topsep=1mm]

\item[\textbf{G1.}]
\textbf{Aggregating activations by counting top activated channels.}
Given the activations for an image, we can view them channel-wise, that is, a collection of 2D matrices where each encodes the magnitude of a detected feature by that channel's learned filter.
We aim to identify which channels have the strongest activation for a given image, so that we can record only the topmost activated channels for every image, and visualize which channels, in aggregate, are most commonly firing a strong activation \textbf{(C1)}. 
This data could then be viewed as a feature of vector for each class, where the features are the counts of images that had a specific channel as a top channel (\autoref{subsec:activation}).

\item[\textbf{G2.}]
\textbf{Aggregating influences by counting previous top influential channels.}
We aim to identify the most influential paths data takes throughout a network.
If aggregated for every image, we could use intermediate outputs of the fundamental convolutional operation used inside of CNNs \textbf{(C2)} to help us determine which channels in a previous layer have the most impact on future channels for a given class of images (\autoref{subsec:influence}).

\item[\textbf{G3.}]
\textbf{Finding what neural networks look for, and how they interact.}
To visualize how low-level concepts near early layers of a network combine to form high-level concepts towards later layers, we seek to form a graph from the entire neural network, using the aggregated influences as an edge list and aggregated activations as vertex values.
With a graph representation, we could leverage the abundant research in graph algorithms, such as Personalized PageRank, to extract a subgraph that best captures the important vertices (neural network channels) and edges (influential paths) in the network (\autoref{subsec:combine}).
\Ags would then describe the most activated channels and attributed paths between channels that ultimately lead the network to a final prediction \textbf{(C3)}.

\item[\textbf{G4.}]
\textbf{Interactive interface to visualize classes \ags of a model.}
We aim to design and develop an interactive interface that can visualize entire \ags (\autoref{sec:ui}).
Our goal is to support users to freely inspect any class within a large neural network classifier to understand what features are learned and how they relate to one another to make predictions for any class \textbf{(C4)}.
Here, we also want to use state-of-the-art deep learning visualization techniques, such as pairing feature visualization with dataset examples, to make channels more interpretable (\autoref{subsec:graph}).

\item[\textbf{G5.}]
\textbf{Deployment using cross-platform, lightweight web technologies.}
To develop a visualization that is accessible for users without specialized computational resources, in \system we use modern web browsers to visualize \ags (\autoref{sec:ui}).
We also open-source our code to support reproducible research \textbf{(C5)}.

\end{itemize}

\section{Model Choice and Background}
\label{sec:model}

\begin{figure*}[t]
 \centering
 \includegraphics[width=0.87\textwidth]{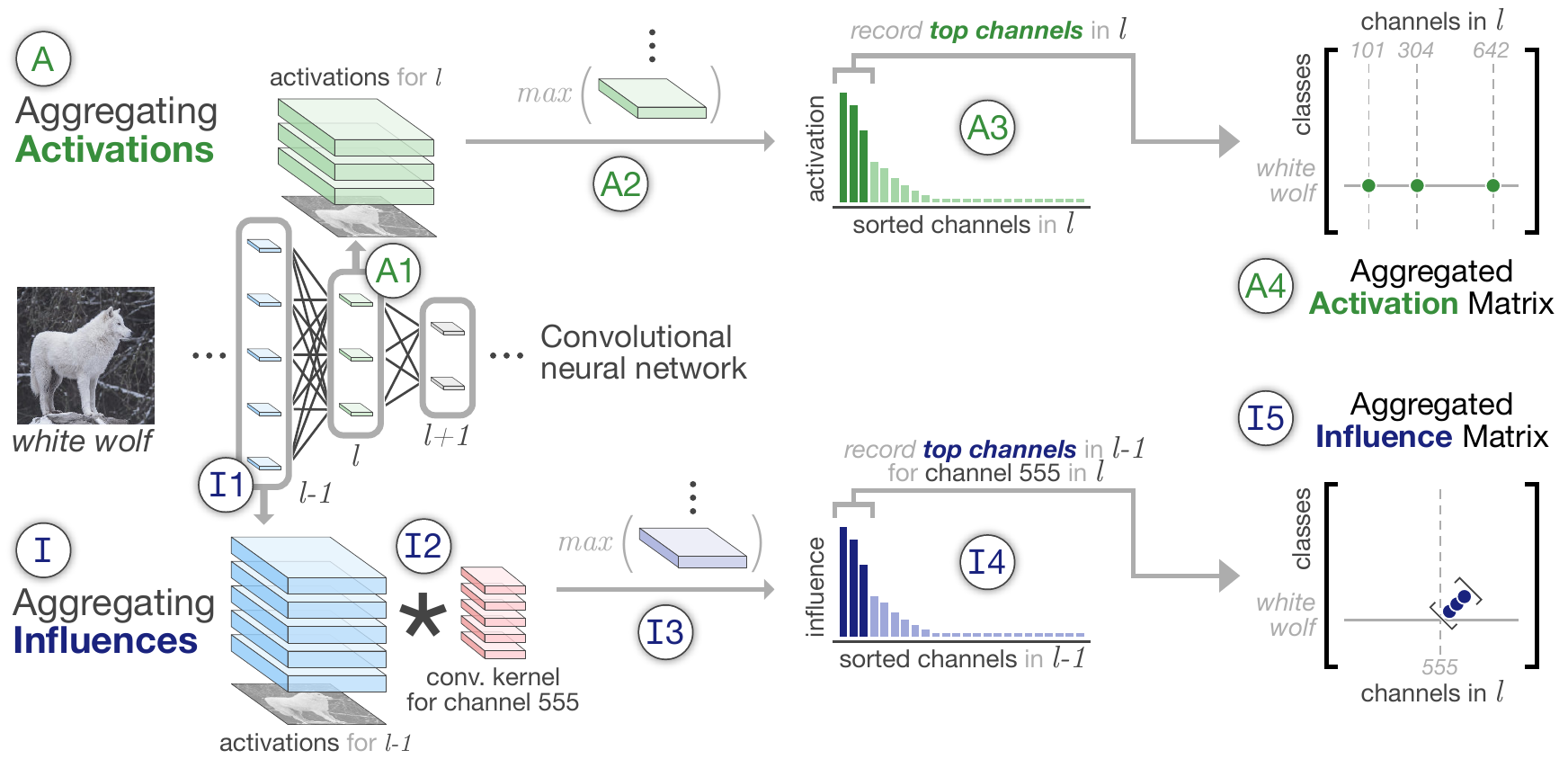}
 \vspace{-3mm}
 \caption{
 Our approach for aggregating activations and influences for a layer $l$.
 \textbf{Aggregating \textcolor{techGreen}{Activations}}: \actStep{A1} given activations at layer $l$, \actStep{A2} compute the max of each 2D channel, and \actStep{A3} record the top activated channels into an \actStep{A4} aggregated activation matrix, which tells us which channels in a layer most activate and represent every class in the model.
 \textbf{Aggregating \textcolor{techPurple}{Influences}}: \infStep{I1} given activations at layer $l-1$, \infStep{I2} convolve them with a convolutional kernel from layer $l$, \infStep{I3} compute the max of each resulting 2D activation map, and \infStep{I4} record the top most influential channels from layer $l-1$ that impact channels in layer $l$ into an \infStep{I5} aggregated influence matrix, which tells us which channels in the previous layer most influence a particular channel in the next layer.
 }
 \vspace{-5mm}
 \label{fig:technique}
\end{figure*}

In this work, we demonstrate our approach on \model~\cite{szegedy2015going}, a prevalent, large-scale convolutional neural network (CNN) that achieves top-5 accuracy of 89.5\% on the ImageNet dataset that contains over 1.2 millions images across 1000 classes.
\model is composed of multiple inception modules: self-contained groups of parallel convolutional layers.
The last layer of each inception module is given a name of the form ``mixed\{number\}\{letter\},'' where the \{number\} and \{letter\} denote the location of a layer in the network; for example, mixed3b (an earlier layer) or mixed4e (a later layer).
In \model, there are 9 such layers: mixed3\{a,b\}, mixed4\{a,b,c,d,e\}, and mixed5\{a,b\}.
While there are more technical complexities regarding neural network design within each inception module, we follow existing interpretability literature and consider the 9 mixed layers as the primary layers of the network~\cite{olah2017feature, olah2018building}.
Although our work makes this model choice, our proposed summarization and visualization techniques can be applied to other neural network architectures in other domains.

\section{Creating \AGs by Aggregation}
\label{sec:technique}

\system introduces two new scalable summarization techniques: 
(1) \textit{\acta} discovers important neurons, and 
(2) \textit{\infa} identifies relationships among such neurons. 
\system combines these techniques to create the novel \textit{\ag} that reveals and summarizes crucial neuron associations and substructures that contribute to a model's outcomes.
\Ags tell us \textit{what} features a neural network detects, and \textit{how} those features are related.
Below, we formulate each technique, and describe how we combine them to generate \ags (\autoref{subsec:combine}) for CNNs.

\subsection{Aggregating Neural Network Activations}
\label{subsec:activation}

We want to understand \textit{what} a neural network is detecting in a dataset.
We propose summarizing how an image dataset is represented throughout a CNN by aggregating individual image \textbf{activations} at each channel in the network, over all of the images in a given class.
This aggregation results in a matrix, $A^l$ for each layer $l$ in a network, where an entry $A^l_{cj}$ roughly represents how \textit{important} channel $j$ (from the $l^{th}$ layer) is for representing images from class $c$.
This measure of importance can be defined in multiple ways, which we discuss formally below.

A convolutional layer contains $C_l$ image kernels (parameters) that are convolved with an input image, $X$, to produce an output image, $Y$, that contains $C_l$ corresponding channels.
For simplicity, we assume that the hyperparameters of the convolutional layer are such that $X$ and $Y$ will have the same height $H$ and width $W$, i.e., $X \in \mathbb{R}^{H \times W \times C_{l-1}}$ and $Y \in \mathbb{R}^{H \times W \times C_l}$.
Each channel in $Y$ is a matrix of values that represent how strongly the corresponding kernel \textit{activated} in each spatial position.
For example, an edge detector kernel will produce a channel, also called an activation map, that has larger values at locations where an edge is present in the input image.
As kernels in convolutional layers are learned during model training, they identify different features that discriminate between different image classes.
It is commonly thought that CNNs build hierarchical feature representations of input images, learning simple edge and shape detectors in early layers of the network, which are combined to form texture detectors, and finally relevant object detectors in later layers of the network~\cite{zeiler2014visualizing} (see \autoref{fig:background}).

A decision must be made on how to aggregate activations over spatial locations in a channel and aggregate activations over all images in a given class.
Ultimately, we want to determine channel importance in a CNN's representation of a class.
As channels roughly represent concepts, we choose the maximum value of a channel as an indicator of how strongly a concept is present, instead of other functions, such as mean, that may dampen the magnitude of relevant channels.

\smallskip
\noindent Alongside \autoref{fig:technique}, our method for aggregation is as follows:
\begin{itemize}[topsep=0mm, itemsep=0mm, parsep=1mm, leftmargin=3mm]

    \item \textbf{Compute activation channel maximums for all images.}
    For each image, \actStep{A1} obtain its activations for a given layer $l$ and \actStep{A2} compute the maximum value per channel.
    This is equivalent to performing Global Max-pooling at each layer in the network.
    Now for each layer, we will have a matrix $Z^l$, where an entry $Z^l_{ij}$ represents the maximum activation of image $i$ over the $j^{th}$ channel in layer $l$. 
    
    \item \textbf{Filter by a particular class.}
    We consider all rows of $Z^l$ whose images belong to the same class, and want to aggregate the maximum activations from these rows to determine which channels are important for detecting the class.
    
    \item \textbf{Aggregation Method 1: taking top $k_{M1}$ channels.}
    For each row, we set the top $k_{M1}$ largest elements to $1$ and others to $0$, then sum over rows.
    Performing this operation for each class in our dataset will result in a matrix $A^l$ from above where an entry $A^l_{cj}$ is the count of the number of times that the $j^{th}$ channel is one of the top $k_{M1}$ channels by maximum activation for all images in class $c$.
    This method ignores the actual maximum activation values, so it will not properly handle cases where a single channel activates strongly for images of a given class (as it will consider $k_{M1}-1$ other channels), or cases where many channels are similarly activated over images of a given class (as it will \textit{only} consider $k_{M1}$ channels as ``important'').
    This observation motivates our second method.
    
    \item \textbf{Aggregation Method 2: taking top $k_{M2}\%$ of channels by weight.} We first scale rows of $Z^l$ to sum to $1$ by dividing by the row sums, $Z^{\prime l}_{ij} = \frac{Z^l_{ij}}{\sum_{n=1}^{N} Z^l_{nj}}$, where $N$ is the number of images.
    Instead of setting the top $k_{M2}$ elements to $1$, as in \textbf{Method 1}, we set the $m$ largest elements of each row to $1$ and the remaining to $0$.
    Here, $m$ is the largest index such that  $\sum_{j \in \text{sorted } Z^{\prime l}_i}^m Z^{\prime l}_{ij} \leq k_{M2}$, where $k_{M2}$ is some small percentage.
    In words, this method first sorts all channels by their maximum activations, then records channels, starting from the largest activated, until the cumulative sum of probability weight from the recorded channels exceeds the threshold.
    Contrary to \textbf{Method 1}, this method adaptively chooses channels that are important for representing a given image, producing a better final class representation.
    
\end{itemize}

\smallskip
\noindent Empirically, we noticed the histograms of max channel activations was often power law distributed, therefore we use \textbf{Method 2} to \actStep{A3} record the top $k_{M2}=3\%$ of channels to include in the \actStep{A4} \textbf{Aggregated \textcolor{techGreen}{Activations}} matrix $A^l$.
In terms of runtime, this process requires only a forward pass through the network.

\subsection{Aggregating Inter-layer Influences}
\label{subsec:influence}

Aggregating activations at each convolutional layer in a network will only give a local description of which channels are important for each class, i.e., from examining $A^l$ we will not know \textit{how} certain channels come to be the most representative for a given class.
Thus, we need a way to calculate how the activations from the channels of a previous layer, $l-1$, \textbf{influence} the activations at the current layer, $l$.
In dense layers, this influence is trivial to compute: the activation at a neuron in $l$ is computed as the weighted sum of activations from neurons in $l-1$.
The influence of a single neuron from $l-1$ is then proportional to the activation of that neuron multiplied by the associated weight to the neuron being examined from $l$. 
In convolutional layers, calculating this influence is more complicated: the activations at a channel in $l$ are computed as the 3D convolution of all of the channels from $l-1$ with a learned kernel tensor.
This operation can be broken down (shown formally later in this section) as a summation of the 2D convolutions of each channel in $l-1$ with a corresponding slice of the appropriate kernel.
The summations of 2D convolutions are similar in structure to the weighted-summations performed by dense layers, however the corresponding ``influence'' of a single channel from $l-1$ on the output of a particular channel in $l$ is a 2D feature map.
We can summarize this feature map into a scalar influence value by using any type of reduce operation, which we discuss further below.

We propose a method for (1) quantifying the \textit{influence} a channel from a previous layer has on the activations of a channel in a following layer, and (2) aggregating influences into a tensor, $I^l$, that can be interpreted similarly to the $A^l$ matrix from the previous section.
Formally, we want to create a tensor $I^l$ for every layer $l$ in a network, where an entry $I^{l}_{cij}$ represents how important channel $i$ from layer $l-1$ is in determining the output of channel $j$ in layer $l$, for all images in class $c$.

First, using the notation from the previous section, we consider how a single channel of $Y$ is created from the channels of $X$.
Let $K^{(j)} \in \mathbb{R}^{H \times W \times C_{l-1}}$ be the $j^{th}$ kernel of our convolutional layer.
Now the operation of a convolutional layer can be written as:
\begin{equation}
\setlength{\abovedisplayskip}{4pt}
\setlength{\belowdisplayskip}{7pt}
    Y_{:,:,j} 
        = \underbrace{ \vphantom{\sum_{i}^{C}}  X \ast K^{(j)}}_{\text{3D convolution}} 
        = \sum_{i=1}^{C_{l-1}} \underbrace{ \vphantom{\sum_{i}^{C}} X_{:,:,i} \ast K^{(j)}_{:,:,i}}_{\text{2D convolution}}
\end{equation}
In words, \infStep{I1} each channel from $X$ is \infStep{I2} convolved with a slice of the $j^{th}$ kernel, and the resulting maps are summed to produce a single channel in $Y$.
We care about the 2D quantity $X_{:,:,i} \ast K^{(j)}_{:,:,i}$ as it contains exactly the contributions of a \textit{single} channel from the previous layer to a channel in the current layer.

Second, we must summarize the quantity $X_{:,:,i} \ast K^{(j)}_{:,:,i}$ into a scalar influence value.
Similarly discussed in \autoref{subsec:activation}, this can be done in many ways, e.g., by summing all values, applying the Frobenius norm, or taking the maximum value.
Each of these summarization methods (i.e., 2D to 1D reduce operations) may lend itself well to exposing interesting connections between channels later in our pipeline.
We chose to \infStep{I3} take the maximum value of $X_{:,:,i} \ast K^{(j)}_{:,:,i}$ as our measure of influence for the image classification task, since this task intuitively considers the largest magnitude of a feature, e.g., how strongly a ``dog ear'' or ``car wheel'' feature is expressed, instead of summing values for example, which might indicate how many places in the image a ``dog ear'' or ``car wheel'' is being expressed.
Also, this mirrors our approach for aggregating activations above.

Lastly, we must aggregate these influence values between channel pairs in consecutive layers, for all images in a given class, i.e., create the proposed $I^{l}$ matrix from the pairwise channel influence values.
This process mirrors the aggregation described previously (\autoref{subsec:activation}), and we follow the same framework.
Let $L^l_{ij}$ be the scalar influence value computed by the previous step \textit{for a single image in class $c$}, between channel $i$ in layer $l-1$ and channel $j$ in layer $l$.
We increment an entry $(c,i,j)$ in the tensor $I^{l}_{cij}$ if $L^l_{ij}$ is one of the top $k_{M1}$ largest values in the column $L^l_{:,j}$ (mirroring \textbf{Method 1} from \autoref{subsec:influence}), or if $L^l_{ij}$ is in the top $k_{M2}\%$ of largest values in $L^l_{:,j}$ (mirroring \textbf{Method 2} (\autoref{subsec:activation}).

Empirically, we noticed the histograms of max influence values were not as often power law distributed as in the previous aggregation of activations, therefore we use \textbf{Method 1} to \infStep{I4} record the top $k_{M1}=5$ channels to include in the \infStep{I5} \textbf{Aggregated \textcolor{techPurple}{Influence}} matrix $I^l$.
Note that \model contains inception modules, groups of branching parallel convolution layers.
Our influence aggregation approach handles these layer depth imbalances by merging paths using the minimum of any two hop edges through an inner layer; this guarantees all edge weights between two hop channels are maximal.
In terms of runtime, this process is more computationally expensive than aggregating activations, since we have to compute all intermediate 2D activation maps; however, with a standard GPU equipped machine is sufficient.
We discuss our experimental setup later in \autoref{subsec:system-design}.

\subsection{Combining Aggregated Activations and Influences to Generate \AGs}
\label{subsec:combine}

Given the aggregated activations $A^l$ and aggregated influences $I^l$ we aim to combine them into a single entity that describes both \textit{what} features a neural network is detecting and \textit{how} those features are related.
We call these \textbf{\ags}, and we describe their generation below.

In essence, neural networks are directed acyclic graphs: they take input data, compute transformations of that data at sequential layers in the network, and ultimately produce an output.
We can leverage this graph structure for our desired representation.
Whereas a common network graph has vertices and connecting edges, our vertices will be the channels of a network (for all layers of the network), and edges connect channels if the channel in the previous layer has a strong influence to a channel in an later, adjacent layer.

\textbf{Using graph algorithms for neural network interpretability.}
Consider the aggregated influences $I^l$ as an edge list; therefore, we can build an ``entire graph'' of a neural network, where edges encode if an image had a path from one channel to another as a top influential path, and the weight of an edge is a count of the number of images for a given class with that path as a top influential path.
Now, for a given class, we want to extract the subgraph that best captures the important vertices (channels) and edges (influential paths) in the network.
Since we have instantiated a typical network graph, we can now leverage the abundant research in graph algorithms.
A natural fit for our task is the Personalized PageRank algorithm~\cite{langville2005survey, page1999pagerank}, which scores each vertex's importance in a graph, 
based on both the graph structure and the weights associated with the graph's vertices and edges.
Specifically, \system operates on the graph produced from all the images of a given class; the algorithm is initialized by and incorporates both vertex information (aggregated activations $A^l$) and edge information (aggregated influences $I^l$) to find a subgraph most relevant for all the provided images.
We normalize each layer's personalization from $A^l$ by dividing by max $A^l$ value for each layer $l$ so that each layer has a PageRank personalization within 0 to 1.
This is required since each layer has a different total number of possible connections (e.g., the first and last layers, mixed3a and mixed5b, only have one adjacent layer, therefore their PageRank values would be biased small).
In summary, we make the full graph of a neural network where vertices are channels from all layers in the network with a personalization from $A^l$, and edges are influences with weights from $I^l$.

\textbf{Extracting \ags.}
After running Personalized PageRank for 100 iterations, the last task is to select vertices based on their computed PageRank values to extract an \ag.
There are many different ways to do this; below we detail our approach.
We first compute histograms of the PageRank vertex values for each layer.
Next, we use the methodology described in \autoref{subsec:activation} for \textbf{Method 2}, where we continue picking vertices with the largest PageRank value until we have reached $k_{M2}\%$ weight for each layer independently.
Empirically, here we set $k_{M2}=7.5\%$ after observing that the PageRank value histograms are roughly power law, indicating that there are only a handful of channels determined important.
Regarding the runtime, the only relevant computation is running PageRank on the full neural network graph, which typically has a few thousands vertices and a few hundred thousand edges.
Using the Python NetworkX\footnote{NetworkX: \url{https://networkx.github.io/}} implementation~\cite{page1999pagerank, langville2005survey}, Personalized PageRank runs in $\sim 30$ seconds for each class.

\section{The \system User Interface}
\label{sec:ui}

From our design goals in \autoref{sec:goals} and our aggregation methodology in \autoref{sec:technique}, we present \system, an interactive system for scalable summarization and interpretation for exploring entire learned classes in large-scale image classifier models (\autoref{fig:teaser}).

The header of \system displays metadata about the visualized image classifier, such as the model and dataset name, the number of classes, and the total number data instances within the dataset. 
As described in \autoref{sec:model}, here we are using \model trained on the 1.2 million image dataset ImageNet that contains 1000 classes.
Beyond the header, the \system user interface is composed of three main interactive views: the \embedding, the \sidebar, and the \graph.
The following section details the representation and features of each view and how they tightly interact with one another. 

\subsection{\embedding: Learned Class Overview}
\label{subsec:embedding}

The first view of \system is the \embedding, a dimensionality reduction overview of all the classes in a model (\autoref{fig:teaser}A).
Given some layer $l$'s $A^l$ matrix, recall an entry in this matrix corresponds to the number of images from one class (row) that had one channel (column) as a top channel.
We can consider \A as a feature matrix for each class where the number of channels in a layer corresponds to the number of features.
For reduction and visualization, the \embedding uses UMAP: a non-linear dimensionality reduction that better preserves global data structure, compared to other techniques like t-SNE, and often provides a better ``big picture'' view of high-dimensional data while preserving local neighbor relations~\cite{mcinnes2018umap}.
Each dot corresponds to one class of the model, with spatial position encoding their similarity.
To explore this embedding, users can freely zoom and pan in the view, and when a user zooms in close enough, labels appear to describe each class (point) so users can easily see how classes within the model compare.
Clicking on a point in the \embedding will update the selection for the remaining views of \system, as described below.

\begin{wrapfigure}{r}{0.25\textwidth}
  \vspace{-2em}
  \begin{center}
    \includegraphics[width=0.25\textwidth]{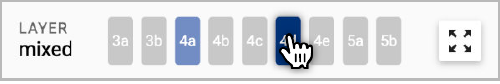}
  \end{center}
  \vspace{-2em}
  \caption{Selectable network minimap animates the \embedding.}
  \vspace{-1em}
  \label{fig:network}
\end{wrapfigure}

\textbf{Selectable neural network minimap.}
At the top of the \embedding sits a small visual representation of the considered neural network; in this case, \model's primary mixed layers are shown (\autoref{fig:network}). 
Since we obtain one $A^l$ matrix for every layer $l$ in the model, to see how the classes related to one another at different layer depths within the network, users can click on one of the other layers to animate the \embedding.
This is useful for obtaining model debugging hints and observing at a high-level how classes are represented throughout a network's layers.

\subsection{\sidebar: Searching and Sorting Classes}
\label{subsec:sidebar}

\begin{wrapfigure}{r}{0.275\textwidth}
\vspace{-2em}
  \begin{center}
    \includegraphics[width=0.275\textwidth]{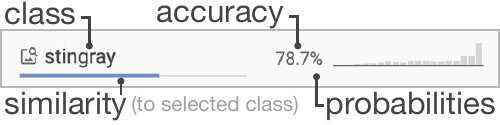}
  \end{center}
  \vspace{-2em}
  \caption{\sidebar visual encoding.}
  \vspace{-1em}
  \label{fig:sidebar-class}
\end{wrapfigure}

Underneath the \embedding sits the \sidebar (\autoref{fig:teaser}B): a scrollable list of all the class of the model, containing high-level class performance statistics.
The first class at the top of the list is the selected class, whose \ag is shown in the \graph, to be discussed in the next section.
The \sidebar is sorted by the similarity of the selected class to all other classes in the model.
For the similarity metric, we compute the cosine similarity using the values from $\A^l$.
Each class is represented as a horizontal bar that contains the class's name, a purple colored bar that indicates its similarity to the selected class (longer purple bars indicate similar classes, and vice versa), the class's top-1 accuracy for classification, and a small histogram of all the images' predicted probabilities within that class (i.e., the output probabilities from the final layer) (\autoref{fig:sidebar-class}).
From this small histogram, users can quickly see how well a class performs.
For example, classes with power law histograms indicate high accuracy, whereas classes with normal distribution histograms indicate underperformance.
Users can then hypothesize whether a model may be biasing particular classes over others, or if underperforming classes have problems with their raw data.

\textbf{Scrolling for context.}
To see where a particular class in the sidebar is located in the \embedding, users can hover over a class to highlight its point and label the \embedding above (\autoref{fig:teaser}A-B).
Since the \sidebar is sorted by class similarity, to see where similar classes lie compared to the selected class, all classes in the \sidebar visible to the user (more technically, in the viewbox of the interface) are also highlighted in the \embedding (\autoref{fig:teaser}A-B).
Scrolling then enables users to quickly see where classes in the \sidebar lie in the \embedding as classes become less similar to the originally selected class to visualize.

\textbf{Sorting and selecting classes.}
To select a new class to visualize, users can click on any class in the \sidebar to update the interface, including resorting the \sidebar by similarity based on the newly selected class and visualize the new class's \ag in the \graph.
Users can also use the search bar to directly search for a known class instead of freely browsing the \sidebar and \embedding.
Lastly, the \sidebar has two additional sorting criteria.
Users can sort the \sidebar by the accuracy, either ascending or descending, to see which classes in the model have the highest and lowest predicted accuracy, providing a direct mechanism to begin to inspect and debug underperforming classes. 

\subsection{\graph: Visual Class Summarization}
\label{subsec:graph}

The \graph is the main view of \system (\autoref{fig:teaser}C).
A small header on top displays some information about the class, similar to that in the \sidebar, and contains a few controls for interacting with the \ag, to be described later.

\textbf{Visualizing \ags.}
Recall from \autoref{subsec:combine} that an \ag is a subgraph of the entire neural network, where the vertices correspond to a class's important channels within a layer, and the edges connect channels based on their influence from the convolution operation.
Our graph visualization design draws inspiration from recent visualization works, such as CNNVis~\cite{liu2017towards}, AEVis~\cite{liu2018analyzing}, and Building Blocks~\cite{olah2018building}, that have successfully leveraged graph based representations for deep learning interpretability.
In the main view of \system, an \ag is shown in a  zoomable and panable canvas that visualizes the graph vertically, where the top corresponds to the last mixed network layer in the network, mixed5b, and the bottom layer corresponds to the first mixed layer, mixed3a (\autoref{fig:teaser}C). 
In essence, the \ag is a directed network with vertices and edges; in \system, we replace vertices with the corresponding channel's feature visualization. 
Each layer, denoted by a label, is a horizontal row of feature visualizations of the \ag.
Each feature visualization is scaled by its magnitude of the number of images within that class that had that channel as a top channel in their prediction, i.e., the value from $A^l$.
Edges are drawn connecting each channel to visualize the important paths data takes during prediction.
Edge thickness is encoded by the influence from one channel to another, i.e., the value from $I^l$.

\textbf{Understanding \ag structure.}
This novel visualization reveals a number of interesting characteristics about how classes behave inside a model.
First, it shows how neural networks build up high-level concepts from low-level features, for example, in the \textit{\textbf{white wolf}} class, early layers learn fur textures, ear detectors, and eye detectors, which all contribute to form face and body detectors in later layers.
Second, the number of visualized channels per layer roughly indicates how many features are needed to represent that class within the network.
For example, in layer mixed5a, the \textit{\textbf{strawberry}} class only has a few large channels, indicating this layer has learned specific object detectors for strawberries already, whereas in the same layer, the \textit{\textbf{drum}} class has many smaller channels, indicating that this layer requires the combination of multiple object detectors working together to represent the class.
Third, users can also see the overall structure of the \ag, and how a model has very few important channels in earlier layers, but as the the network progress, certain channels grow in size and begin to learn high-level features about what an image contains.

\textbf{Inspecting channels and connections in \ags.}
Besides displaying the feature visualization at each vertex, there are a number of different complementary data that is visualized to help interpret what a model has learned for a given class \ag.
It has been shown that for interpreting channels in a neural network, feature visualization is not always enough~\cite{olah2017feature}; however, displaying example image patches from the entire dataset next to a feature visualization helps people better understand what the channel is detecting.
We apply a similar approach, where hovering over a channel reveals 10 image patches from the entire dataset that most maximize this specific channel (\autoref{fig:teaser}C).
Pairing feature visualization with dataset examples helps understand what the channel is detecting in the case where a feature visualization alone is hard to decipher.
When a user hovers over a channel, \system also highlights the edges that flow in and out of that specific channel by coloring the edges and animating them within the \ag. 
This is helpful for understanding which and how much channels in a previous layer contribute to a new channel in a later layer.
Users can also hover over the edges of an \ag to color and animate that specific edge and its endpoint channels, similar to the interaction used when hovering over channels.
Lastly, users can get more insight into what feature a specific channel has learned by hovering left to right on a channel to see the feature visualization change to display four other feature visualizations generated with \textit{diversity}: a technique used to create multiple feature visualizations for a specific channel at once that reveals different areas of latent space that a channel has learned~\cite{olah2017feature}.
This interaction is inspired from commercial photo management applications where users can simply hover over an image album's thumbnail to quickly preview what images are are inside.

\textbf{Dynamic drill down and filtering.}
When exploring an \ag, users can freely zoom and pan the entire canvas, and return to the zoomed-out overview of the visualization via a button included in the options bar above the \ag.
In the case of a large \ag where there are too many channels and edges, in the options bar there is a slider that when dragged, filters the the channels of the \ag by their importance from $\A^l$.
This interaction technique draws inspiration from existing degree-of-interest graph exploration research, where users can dynamically filter and highlight a subset of the most important channels (vertices) and connections (edges) based on computed scores~\cite{furnas1986generalized, van2009search, crnovrsanin2011visual, kairam2015refinery}.
Dragging the slider triggers an animation where the filtered-out channels and their edges are removed from the \ag, and the remaining visualization centers itself for each layer.
With the additional width and height sliders, these interactions add dynamism to the \ag, where it fluidly animates and updates to users deciding the scale of the visualization.

\subsection{System Design}
\label{subsec:system-design}
To broaden access to our work, \system is web-based and can be accessed from any modern web-browser.
\system uses the standard HTML/CSS/JavaScript stack, and D3.js\footnote{D3.js: \url{https://d3js.org/}} for rendering SVGs.
We ran all our deep learning code on a NVIDIA DGX 1, a workstation with 8 GPUs, with 32GB of RAM each, 80 CPU cores, and 504GB of RAM.
With this machine we could generate everything required for \textit{all 1000 ImageNet} classes---aggregating activations, aggregating influences, and combining them with PageRank (implementation from NetworkX) to form \ags---and perform post-processing under 24 hours.
However, visualizing a single class on one GPU takes only a few minutes.
The \textit{Lucid} library is used for creating feature visualizations\footnote{Lucid: \url{https://github.com/tensorflow/lucid}}, and dataset examples are used from the appendix\footnote{\url{https://github.com/distillpub/post--feature-visualization} } of~\cite{olah2017feature}.

\section{Neural Network Exploration Scenarios}
\label{sec:results}

\subsection{Unexpected Semantics Within a Class}
A problem with deploying neural networks in critical domains is their lack of interpretability, specifically, can model developers be confident that their network has learned what they think it has learned?
We can answer perplexing questions like these with \system.
For example, in \autoref{fig:teaser}, consider the \textbf{\textit{tench}} class (a type of yellow-brown fish).
Starting from the first layer, as we explore the \ag for \textbf{\textit{tench}} we notice there are no fish or water features, but there are many ``finger'', ``hand'', and ``people'' detectors.
It is not until a middle layer, mixed4d, that the first fish and scale detectors are seen (\autoref{fig:teaser}C, callout); however, even these detectors focus solely on the body of the fish (there is no fish eye, face, or fin detectors).
Inspecting dataset examples reveals many image patches where we see people's fingers holding fish, presumably after catching them.
This prompted us to inspect the raw data for the \textbf{\textit{tench}} class, where indeed, most of the images are of a person holding the fish.
We conclude that, unexpectedly, the model uses people detectors and in combination with brown fish body and scale detectors to represent the \textbf{\textit{tench}} class.
Generally, we would not expect ``people'' as an essential feature for classifying fish.

\begin{figure}[t]
 \centering
 \includegraphics[width=0.99\columnwidth]{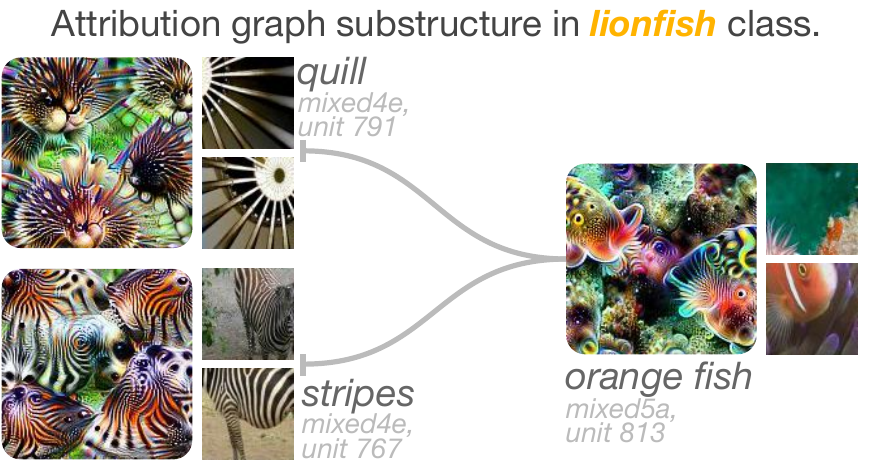}
 \vspace{-3mm}
 \caption{
 An example substructure from the \textit{\textbf{\textcolor{unexOrange}{lionfish}}} \ag that shows unexpected texture features, like ``quills'' and ``stripes,'' influencing top activated channels for a final layer's ``orange fish'' feature (some \textit{\textbf{\textcolor{unexOrange}{lion fish}}} are reddish-orange, and have white fin rays).
 }
 \label{fig:lionfish}
 \vspace{-5mm}
\end{figure}

This surprising finding motivated us to seek another class of fish that people do not normally hold to compare against, such as a \textit{\textbf{\textcolor{unexOrange}{lionfish}}} (due to their venomous spiky fin rays).
Visualizing the \textit{\textbf{\textcolor{unexOrange}{lionfish}}} \ag confirms our suspicion (\autoref{fig:lionfish}): there are not any people object detectors in its \ag.
However, we discover yet another unexpected combination of features: there are few fish part detectors while there are many texture features, e.g., stripes and quills.
It is not until the final layers of the network where a highly activated channel detects orange fish in water, which uses the stripe and quill detectors.
Therefore we deduce that the \textit{\textbf{\textcolor{unexOrange}{lionfish}}} class is composed of a striped body in the water with long, thin quills.
Whereas the \textbf{\textit{tench}} had unexpected people features, the \textit{\textbf{\textcolor{unexOrange}{lionfish}}} lacked fish features.
Regardless, findings such as these can help people more confidently deploy models when they know what composition of features results in a prediction.

\subsection{Mixed Class Association Throughout Layers}
While inspecting the \embedding, we noticed some classes' embedding positions shift greatly between adjacent layers.
This cross-layer embedding comparison is possible since each layer's embedding uses the previous layer's embedding as an initialization.
Upon inspection, the classes that changed the most were classes that were either a combination of existing classes or had \textit{mixed primary associations}.

\begin{figure}[tb]
 \centering
 \includegraphics[width=0.98\columnwidth]{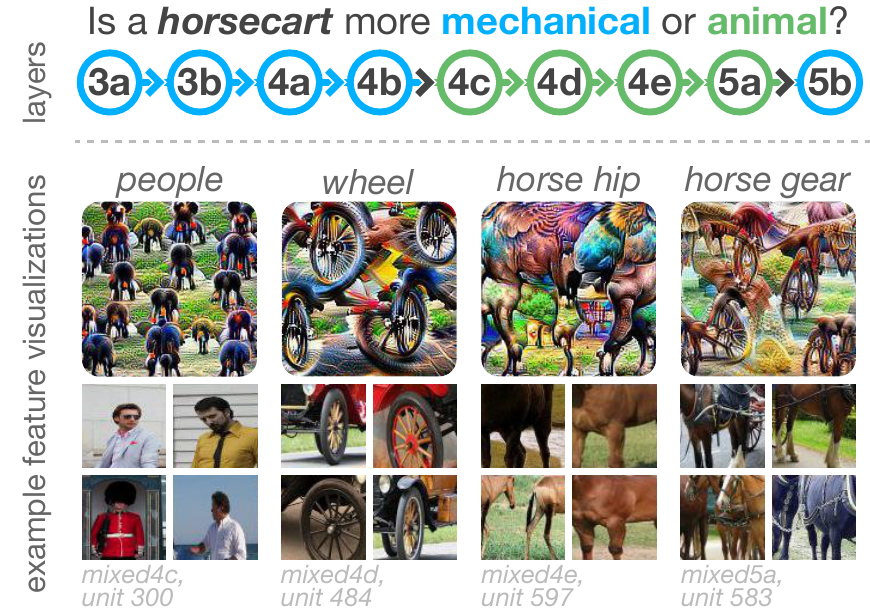}
\vspace{-3mm}
 \caption{
 Using \system we can find classes with mixed semantics that shift their primary associations throughout the network layers.
 For example, early in the network, \textbf{\textit{horsecart}} is most similar to  \textbf{\textcolor{mixedBlue}{mechanical}} classes (e.g., harvester, thresher,  snowplow), towards the middle it shifts to be nearer to  \textbf{\textcolor{mixedGreen}{animal}} classes (e.g., bison, wild boar,  ox), but ultimately returns to have a stronger \textbf{\textcolor{mixedBlue}{mechanical}} association at the network output.
 }
 \label{fig:mixed-class}
 \vspace{-5mm}
\end{figure}

For example, consider the \textbf{\textit{horsecart}} class.
For each layer, we can inspect the nearest neighbors of \textbf{\textit{horsecart}} to check its similarity to other classes.
We find that \textbf{\textit{horsecart}} in the early layers is similar to other \textbf{\textcolor{mixedBlue}{mechanical}} classes, e.g., harvester, thresher, and snowplow.
This association shifts in the middle layers where \textbf{\textit{horsecart}} moves to be near \textbf{\textcolor{mixedGreen}{animal}} classes, e.g., bison, wild boar, and ox.
However, \textbf{\textit{horsecart}} flips back at the final convolutional layer, 
returning to a \textbf{\textcolor{mixedBlue}{mechanical}} association (\autoref{fig:mixed-class}, top).
To better understand what features compose a \textbf{\textit{horsecart}}, we inspect its \ag and find multiple features throughout all the layers that contain people, spoke wheels, horse hips, and eventually horse bodies with saddles and mechanical gear (\autoref{fig:mixed-class}, bottom).
Mixed semantic classes like \textbf{\textit{horsecart}} allow us to test if certain classes are semantic combinations of others and probe deeper into understanding how neural networks build hierarchical representations.

\subsection{Discriminable Features in Similar Classes}

\begin{figure*}[t]
 \centering
 \includegraphics[width=0.9\linewidth]{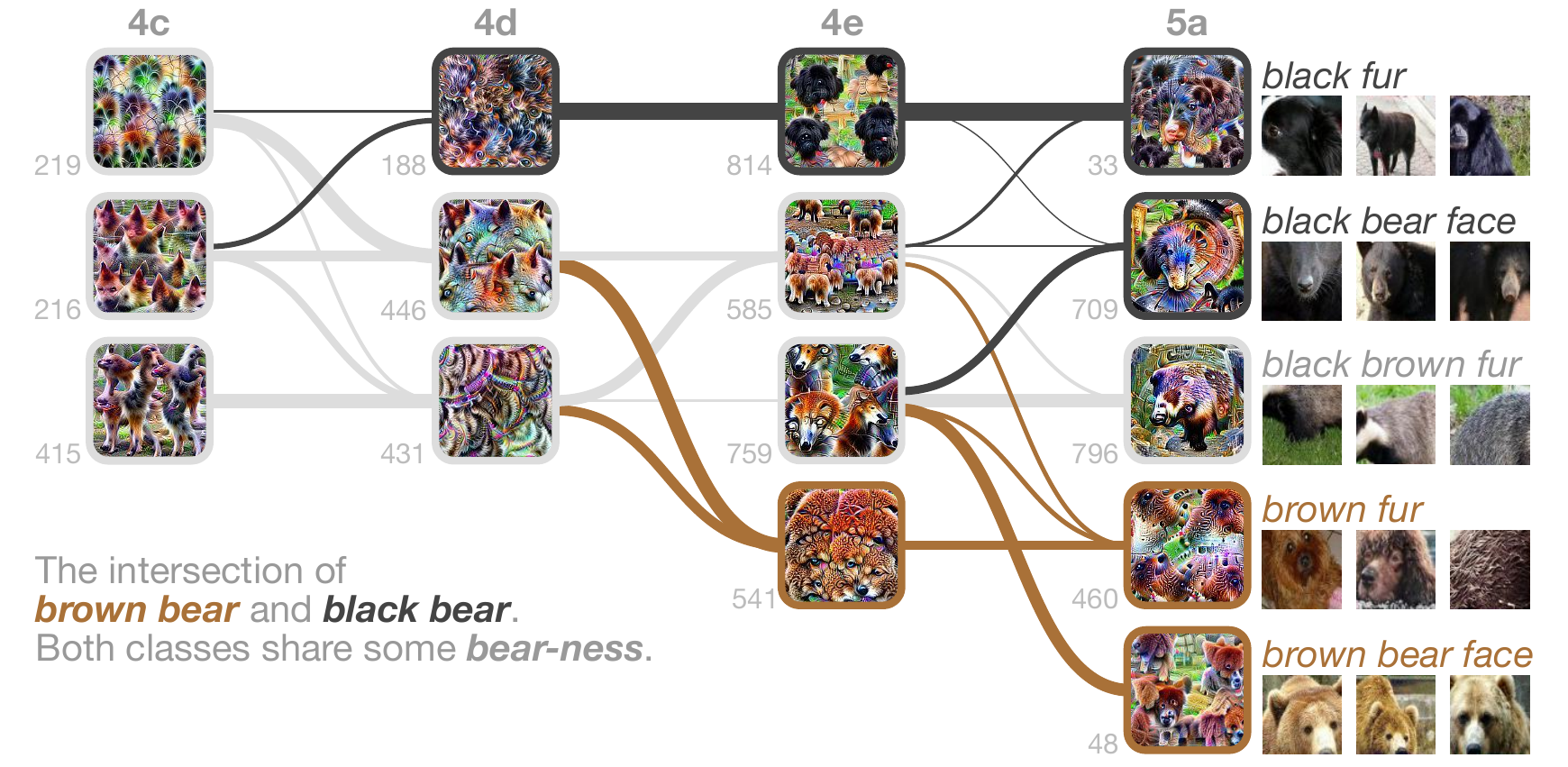}
 \vspace{-3mm}
 \caption{With \ags, we can compare classes throughout layers of a network.
 Here we compare two similar classes: \blackBear and \brownBear.
 From the intersection of their \ags, we see both classes share features related to \bearness, but diverge towards the end of the network using fur color and face color as discriminable features.
 This feature discrimination aligns with how humans might classify bears.
 }
 \label{fig:discriminable}
 \vspace{-5mm}
\end{figure*}

Since neural networks are loosely inspired by the human brain, in the broader machine learning literature there is great interest to understand if decision rationale in neural networks is similar to that of humans.
With \ags, we can further  to answer this question by comparing classes throughout layers of a network.

For example, consider the \blackBear and \brownBear classes.
A human would likely say that color is the discriminating difference between these classes.
By taking the \textit{intersection} of their \ags, we can see what features are shared between the classes, as well as any discriminable features and connections.
In \autoref{fig:discriminable}, we see in earlier layers (mixed4c) that both \blackBear and \brownBear share many features, but as we move towards the output, we see multiple diverging paths and channels that distinguish features for each class.
Ultimately, we see individual black and brown fur and bear face detectors, while some channels represent general \bearness.
Therefore, it appears \model classifies \blackBear and \brownBear based on color, which may be the primary feature humans may classify by.
This is only one example, and it is likely that these discriminable features do not always align with what we would expect; however, \ags give us a mechanism to test hypotheses like these.

\subsection{Finding Non-semantic Channels}
\begin{figure}[t!]
 \centering
 \includegraphics[width=0.92\columnwidth]{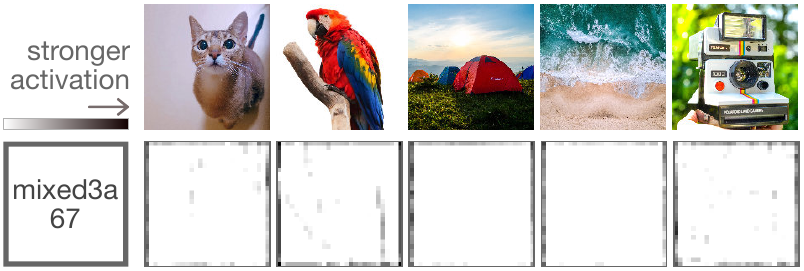}
 \vspace{-2mm}
 \caption{
    Using \system on \model we found non-semantic channels that detect irrelevant features, regardless of the input image, e.g., in layer mixed3a, channel 67 is activated by the frame of an image.
 }
 \label{fig:outlier}
 \vspace{-5mm}
\end{figure}

Using \system, we quickly found several channels that detected non-semantic, irrelevant features, regardless of input image or class (verified manually with 100+ classes, computationally with all).
For example, in layer mixed3a, channel 67 activates to the image frame, as seen in \autoref{fig:outlier}.
We found 5 total non-semantic channels, including mixed3a 67, mixed3a 190, mixed3b 390, mixed3b 399, and mixed3b 412.
Upon finding these, we reran our algorithm for aggregating activations and influences, and generated all \ags with these channels excluded from the computation, since they consistently produced high activation values but were incorrectly indicating important features in many classes.
Although \system leverages recent feature visualization research~\cite{olah2017feature} to visualize channels, it does not provide an automated way to measure the semantic quality of channels.
We point readers to the appendix of~\cite{olah2017feature} to explore this important future research direction.

\subsection{Informing Future Algorithm Design}
We noticed that some classes (e.g., \textit{\textbf{zebra}}, \textit{\textbf{green mamba}}) have only a few important channels in the middle layers of the network, indicating that these channels could have enough information to act as a predictor for the given class.
This observation implies that it may be prudent to make classification decisions at different points in the network, as opposed to after a single softmax layer at the output.
More specifically, per the $\A^l$ matrices, we can easily find these channels (in all layers) that maximally activates for each class.
We could then perform a MaxPooling operation at each of these channels, followed by a Dense layer classifier to form a new ``model'' that only uses the most relevant features for each class to make a decision.

The inspiration for this proposed algorithm is a direct result of the observations made possible by \system.
Furthermore, our proposed methodology makes it easy to test whether the motivating observation holds true for other networks besides \model.
It could be the case that single important channels for certain classes are a result of the training with multiple softmax `heads' used by \model; however, without \system, checking this would be difficult.

\section{Discussion and Future Work}
\label{sec:discussion}

\textbf{Interactive visual comparison of \ags.}
Currently, \system interactively visualizes single \ags.
However, there is great opportunity to support automatically, visual comparison between multiple \ags.
Example comparison operations include computing \ag difference, union, and intersection.

\textbf{Mining \ags for subgraph motifs.}
Since \ags are regular network graphs, we can leverage data mining and graph analysis techniques to find the most common motifs, e.g., all mammal classes may have three specific channels that form a triangle that is always activated highly, or maybe all car classes share only single path throughout the network.
Extracting these smaller subgraph motifs could give deep insight into how neural networks arrange hierarchical concepts inside their internal structure.

\textbf{Visualizing other neural network models.}
We justify our model choice in \autoref{sec:model}, but an immediate avenue for future work explores generating \ags on other CNN models.
Simpler models like VGG~\cite{simonyan2014very} can be easily adapted with our approach, but more complex networks like ResNets~\cite{he2016deep} will require a small modification for computing attribution and influences (e.g., considering skip connections between layers as additional graph edges).
Our approach also may be adopted for exploring neural network components of model architectures that provide activation information (e.g., the two individual networks within a GAN~\cite{goodfellow2014generative}, but not their interaction).

\textbf{Better \ag generation.}
Computing neural network attribution remains an active area of research: there is no consensus of the best way to compute attribution~\cite{olah2018building, selvaraju2017grad, fong2017interpretable, kindermans2017learning, sundararajan2017axiomatic, zeiler2014visualizing, simonyan2013deep}.
To generate \ags, we use activation aggregation as an initialization for personalized PageRank on the entire network from aggregated influences.
While this is one effective way to generate \ags, there could be other ways to generate graph explanations that describe learned neural network representations.
If so, this will only improve the value of \system's visualizations.
For example, layer-wise relevance propagation~\cite{bach2015pixel} could be used to seed our aggregation methods using relevance scores instead of neuron activations.
Conversely, exploring \ags using less-contributing channels could be a novel way to discover non-relevant features.
However, aggregation over spatial positions and instances, a main contribution of \system, will still be necessary given any other measure of neuron importance.

\textbf{Hyperparameter selections.}
Our approach has a few hyperparameters choices, including determining how many channels to record per image when aggregating activations and computing attribution graph influences, as well as what PageRank threshold to set for creating the final visualizations.
However, since our approach was designed to take advantage of data at scale, in our tests we do not see many differences in the limit that the number of images increases.
Note that while our approach benefits from scale, both the aggregation and visualization work on arbitrary dataset sizes, e.g., a single image, hundreds, or thousands.

\textbf{Longitudinal evaluation of impacts in practice.} 
We presented Summit to ML researchers and scientists at industry and government research labs, and discussed plans to conduct long-term studies to test Summit on their own models.
We plan to investigate how Summit may inform algorithmic model design, prompt data collection for ill-represented classes, and discover latent properties of deployed models.

\section{Conclusion}
\label{sec:conclusion}

As deep learning is increasingly used in decision-making tasks, it is important to understand how neural networks learn their internal representations of large datasets.
In this work, we present \system, an interactive system that scalably and systematically summarizes and visualizes what features a deep learning model has learned and how those features interact to make predictions.
The \system visualization runs in modern web browsers and is open-sourced.
We believe our summarization approach that builds entire class representations is an important step for developing higher-level explanations for neural networks. 
We hope our work will inspire deeper engagement from both the information visualization and machine learning communities to further develop human-centered tools for artificial intelligence~\cite{abdul2018trends, olah2017research}.

\acknowledgments{We thank Nilaksh Das, the Georgia Tech Visualization Lab, and the anonymous reviewers for their support and constructive feedback.
This work is supported by a NASA Space Technology Research Fellowship and NSF grants IIS-1563816, CNS-1704701, and TWC-1526254.}

\bibliographystyle{abbrv-doi}

\bibliography{19-summit-vast}

\begin{thebibliography}{10}

\bibitem{abdul2018trends}
A.~Abdul, J.~Vermeulen, D.~Wang, B.~Y. Lim, and M.~Kankanhalli.
\newblock Trends and trajectories for explainable, accountable and intelligible
  systems: An hci research agenda.
\newblock In {\em Proceedings of the 2018 CHI Conference on Human Factors in
  Computing Systems}, p. 582. ACM, 2018.

\bibitem{amershi2015modeltracker}
S.~Amershi, M.~Chickering, S.~M. Drucker, B.~Lee, P.~Simard, and J.~Suh.
\newblock Modeltracker: Redesigning performance analysis tools for machine
  learning.
\newblock In {\em Proceedings of the 33rd Annual ACM Conference on Human
  Factors in Computing Systems}, pp. 337--346. ACM, 2015.

\bibitem{bach2015pixel}
S.~Bach, A.~Binder, G.~Montavon, F.~Klauschen, K.-R. M{\"u}ller, and W.~Samek.
\newblock On pixel-wise explanations for non-linear classifier decisions by
  layer-wise relevance propagation.
\newblock {\em PLOS ONE}, 10(7):e0130140, 2015.

\bibitem{bau2017network}
D.~Bau, B.~Zhou, A.~Khosla, A.~Oliva, and A.~Torralba.
\newblock Network dissection: Quantifying interpretability of deep visual
  representations.
\newblock In {\em Proceedings of the IEEE Conference on Computer Vision and
  Pattern Recognition}, pp. 6541--6549, 2017.

\bibitem{bilal2018convolutional}
A.~Bilal, A.~Jourabloo, M.~Ye, X.~Liu, and L.~Ren.
\newblock Do convolutional neural networks learn class hierarchy?
\newblock {\em IEEE transactions on visualization and computer graphics},
  24(1):152--162, 2018.

\bibitem{biran2017explanation}
O.~Biran and C.~Cotton.
\newblock Explanation and justification in machine learning: A survey.
\newblock In {\em IJCAI Workshop on Explainable AI}, 2017.

\bibitem{carter2019activation}
S.~Carter, Z.~Armstrong, L.~Schubert, I.~Johnson, and C.~Olah.
\newblock Activation atlas.
\newblock {\em Distill}, 4(3):e15, 2019.

\bibitem{crnovrsanin2011visual}
T.~Crnovrsanin, I.~Liao, Y.~Wu, and K.-L. Ma.
\newblock Visual recommendations for network navigation.
\newblock In {\em Computer Graphics Forum}, vol.~30, pp. 1081--1090. Wiley
  Online Library, 2011.

\bibitem{das2018shield}
N.~Das, M.~Shanbhogue, S.-T. Chen, F.~Hohman, S.~Li, L.~Chen, M.~E. Kounavis,
  and D.~H. Chau.
\newblock Shield: Fast, practical defense and vaccination for deep learning
  using jpeg compression.
\newblock In {\em Proceedings of the 24th ACM SIGKDD International Conference
  on Knowledge Discovery \& Data Mining}, pp. 196--204. ACM, 2018.

\bibitem{doshi2017towards}
F.~Doshi-Velez and B.~Kim.
\newblock Towards a rigorous science of interpretable machine learning.
\newblock {\em arXiv preprint arXiv:1702.08608}, 2017.

\bibitem{erhan2009visualizing}
D.~Erhan, Y.~Bengio, A.~Courville, and P.~Vincent.
\newblock Visualizing higher-layer features of a deep network.
\newblock {\em University of Montreal}, 1341:3, 2009.

\bibitem{fong2018net2vec}
R.~Fong and A.~Vedaldi.
\newblock Net2vec: Quantifying and explaining how concepts are encoded by
  filters in deep neural networks.
\newblock In {\em Proceedings of the IEEE Conference on Computer Vision and
  Pattern Recognition}, pp. 8730--8738, 2018.

\bibitem{fong2017interpretable}
R.~C. Fong and A.~Vedaldi.
\newblock Interpretable explanations of black boxes by meaningful perturbation.
\newblock In {\em Proceedings of the IEEE International Conference on Computer
  Vision}, pp. 3429--3437, 2017.

\bibitem{furnas1986generalized}
G.~W. Furnas.
\newblock {\em Generalized fisheye views}, vol.~17.
\newblock Bell Communications Research. Morris Research and Engineering
  Center~…, 1986.

\bibitem{gatys2016image}
L.~A. Gatys, A.~S. Ecker, and M.~Bethge.
\newblock Image style transfer using convolutional neural networks.
\newblock In {\em Proceedings of the IEEE Conference on Computer Vision and
  Pattern Recognition}, pp. 2414--2423, 2016.

\bibitem{gilpin2018explaining}
L.~H. Gilpin, D.~Bau, B.~Z. Yuan, A.~Bajwa, M.~Specter, and L.~Kagal.
\newblock Explaining explanations: An approach to evaluating interpretability
  of machine learning.
\newblock {\em arXiv preprint arXiv:1806.00069}, 2018.

\bibitem{goodfellow2016deep}
I.~Goodfellow, Y.~Bengio, and A.~Courville.
\newblock {\em Deep learning}.
\newblock 2016.

\bibitem{goodfellow2014generative}
I.~Goodfellow, J.~Pouget-Abadie, M.~Mirza, B.~Xu, D.~Warde-Farley, S.~Ozair,
  A.~Courville, and Y.~Bengio.
\newblock Generative adversarial nets.
\newblock In {\em Advances in neural information processing systems}, pp.
  2672--2680, 2014.

\bibitem{harley2015isvc}
A.~W. Harley.
\newblock An interactive node-link visualization of convolutional neural
  networks.
\newblock In {\em ISVC}, pp. 867--877, 2015.

\bibitem{he2016deep}
K.~He, X.~Zhang, S.~Ren, and J.~Sun.
\newblock Deep residual learning for image recognition.
\newblock In {\em Proceedings of the IEEE Conference on Computer Vision and
  Pattern Recognition}, pp. 770--778, 2016.

\bibitem{hohman2019gamut}
F.~Hohman, A.~Head, R.~Caruana, R.~DeLine, and S.~M. Drucker.
\newblock Gamut: A design probe to understand how data scientists understand
  machine learning models.
\newblock In {\em Proceedings of the SIGCHI Conference on Human Factors in
  Computing Systems}. ACM, 2019.

\bibitem{hohman2018visual}
F.~Hohman, M.~Kahng, R.~Pienta, and D.~H. Chau.
\newblock Visual analytics in deep learning: An interrogative survey for the
  next frontiers.
\newblock {\em IEEE Transactions on Visualization and Computer Graphics},
  25(8):2674--2693, Aug 2019. doi: {{%
10\hspace{.1pt}\discretionary{.}{%
}{.}\hspace{.4pt}1109\discretionary{/}{%
}{/}TVCG\hspace{.1pt}\discretionary{.}{%
}{.}\hspace{.4pt}2018\hspace{.1pt}\discretionary{.}{%
}{.}\hspace{.4pt}2843369}}


\bibitem{jean2016combining}
N.~Jean, M.~Burke, M.~Xie, W.~M. Davis, D.~B. Lobell, and S.~Ermon.
\newblock Combining satellite imagery and machine learning to predict poverty.
\newblock {\em Science}, 353(6301):790--794, 2016.

\bibitem{kahng2018activis}
M.~Kahng, P.~Y. Andrews, A.~Kalro, and D.~H.~P. Chau.
\newblock Activis: Visual exploration of industry-scale deep neural network
  models.
\newblock {\em IEEE Transactions on Visualization and Computer Graphics},
  24(1):88--97, 2018.

\bibitem{kahng2019gan}
M.~Kahng, N.~Thorat, D.~H.~P. Chau, F.~B. Vi{\'e}gas, and M.~Wattenberg.
\newblock Gan lab: Understanding complex deep generative models using
  interactive visual experimentation.
\newblock {\em IEEE Transactions on Visualization and Computer Graphics},
  25(1):310--320, 2019.

\bibitem{kairam2015refinery}
S.~Kairam, N.~H. Riche, S.~Drucker, R.~Fernandez, and J.~Heer.
\newblock Refinery: Visual exploration of large, heterogeneous networks through
  associative browsing.
\newblock In {\em Computer Graphics Forum}, vol.~34, pp. 301--310. Wiley Online
  Library, 2015.

\bibitem{kim2017interpretability}
B.~{Kim}, W.~M., J.~{Gilmer}, C.~C., W.~J., , F.~{Viegas}, and R.~{Sayres}.
\newblock {Interpretability Beyond Feature Attribution: Quantitative Testing
  with Concept Activation Vectors (TCAV)}.
\newblock {\em ICML}, 2018.

\bibitem{kindermans2017learning}
P.-J. Kindermans, K.~T. Sch{\"u}tt, M.~Alber, K.-R. M{\"u}ller, D.~Erhan,
  B.~Kim, and S.~D{\"a}hne.
\newblock Learning how to explain neural networks: Patternnet and
  patternattribution.
\newblock {\em arXiv preprint arXiv:1705.05598}, 2017.

\bibitem{langville2005survey}
A.~N. Langville and C.~D. Meyer.
\newblock A survey of eigenvector methods for web information retrieval.
\newblock {\em SIAM Review}, 47(1):135--161, 2005.

\bibitem{lipton2016mythos}
Z.~C. Lipton.
\newblock The mythos of model interpretability.
\newblock {\em ICML Workshop on Human Interpretability in Machine Learning},
  2016.

\bibitem{liu2018analyzing}
M.~Liu, S.~Liu, H.~Su, K.~Cao, and J.~Zhu.
\newblock Analyzing the noise robustness of deep neural networks.
\newblock {\em IEEE Conference on Visual Analytics Science and Technology},
  2018.

\bibitem{liu2017towards}
M.~Liu, J.~Shi, Z.~Li, C.~Li, J.~Zhu, and S.~Liu.
\newblock Towards better analysis of deep convolutional neural networks.
\newblock {\em IEEE Transactions on Visualization and Computer Graphics},
  23(1):91--100, 2017.

\bibitem{liu2018artificial}
Y.~Liu, T.~Kohlberger, M.~Norouzi, G.~Dahl, J.~Smith, A.~Mohtashamian,
  N.~Olson, L.~Peng, J.~Hipp, and M.~Stumpe.
\newblock Artificial intelligence-based breast cancer nodal metastasis
  detection.
\newblock {\em Archives of Pathology \& Laboratory Medicine}, 143(7):859--868,
  2019.

\bibitem{lu2017state}
Y.~Lu, R.~Garcia, B.~Hansen, M.~Gleicher, and R.~Maciejewski.
\newblock The state-of-the-art in predictive visual analytics.
\newblock In {\em Computer Graphics Forum}, vol.~36, pp. 539--562. Wiley Online
  Library, 2017.

\bibitem{mcinnes2018umap}
L.~{McInnes}, J.~{Healy}, and J.~{Melville}.
\newblock {UMAP: Uniform Manifold Approximation and Projection for Dimension
  Reduction}.
\newblock {\em ArXiv e-prints}, Feb. 2018.

\bibitem{montavon2017methods}
G.~Montavon, W.~Samek, and K.-R. M{\"u}ller.
\newblock Methods for interpreting and understanding deep neural nnetworks.
\newblock {\em Digital Signal Processing}, 2017.

\bibitem{mordvintsev2015inceptionism}
A.~Mordvintsev, C.~Olah, and M.~Tyka.
\newblock Inceptionism: Going deeper into neural networks.
\newblock {\em Google Research Blog}, 2015.

\bibitem{nguyen2017plug}
A.~Nguyen, J.~Clune, Y.~Bengio, A.~Dosovitskiy, and J.~Yosinski.
\newblock Plug \& play generative networks: Conditional iterative generation of
  images in latent space.
\newblock In {\em Proceedings of the IEEE Conference on Computer Vision and
  Pattern Recognition}, pp. 4467--4477, 2017.

\bibitem{olah2017research}
C.~Olah and S.~Carter.
\newblock Research debt.
\newblock {\em Distill}, 2017.
\newblock https://distill.pub/2017/research-debt. doi: {{%
10\hspace{.1pt}\discretionary{.}{%
}{.}\hspace{.4pt}23915\discretionary{/}{%
}{/}distill\hspace{.1pt}\discretionary{.}{%
}{.}\hspace{.4pt}00005}}


\bibitem{olah2017feature}
C.~Olah, A.~Mordvintsev, and L.~Schubert.
\newblock Feature visualization.
\newblock {\em Distill}, 2(11):e7, 2017.

\bibitem{olah2018building}
C.~Olah, A.~Satyanarayan, I.~Johnson, S.~Carter, L.~Schubert, K.~Ye, and
  A.~Mordvintsev.
\newblock The building blocks of interpretability.
\newblock {\em Distill}, 3(3):e10, 2018.

\bibitem{page1999pagerank}
L.~Page, S.~Brin, R.~Motwani, and T.~Winograd.
\newblock The pagerank citation ranking: Bringing order to the web.
\newblock Technical report, Stanford InfoLab, 1999.

\bibitem{pan2010survey}
S.~J. Pan and Q.~Yang.
\newblock A survey on transfer learning.
\newblock {\em IEEE Transactions on Knowledge and Data Engineering},
  22(10):1345--1359, 2010.

\bibitem{council2016gdpr}
Parliament and C.~of~the European~Union.
\newblock General data protection regulation.
\newblock 2016.

\bibitem{ren2017squares}
D.~Ren, S.~Amershi, B.~Lee, J.~Suh, and J.~D. Williams.
\newblock Squares: Supporting interactive performance analysis for multiclass
  classifiers.
\newblock {\em IEEE Transactions on Visualization and Computer Graphics},
  23(1):61--70, 2017.

\bibitem{ribeiro2016should}
M.~T. Ribeiro, S.~Singh, and C.~Guestrin.
\newblock Why should i trust you?: Explaining the predictions of any
  classifier.
\newblock In {\em Proceedings of the 22nd ACM SIGKDD International Conference
  on Knowledge Discovery and Data Mining}, pp. 1135--1144. ACM, 2016.

\bibitem{russakovsky2015imagenet}
O.~Russakovsky, J.~Deng, H.~Su, J.~Krause, S.~Satheesh, S.~Ma, Z.~Huang,
  A.~Karpathy, A.~Khosla, M.~Bernstein, et~al.
\newblock Imagenet large scale visual recognition challenge.
\newblock {\em International Journal of Computer Vision}, 115(3):211--252,
  2015.

\bibitem{selvaraju2018choose}
R.~R. Selvaraju, P.~Chattopadhyay, M.~Elhoseiny, T.~Sharma, D.~Batra,
  D.~Parikh, and S.~Lee.
\newblock Choose your neuron: Incorporating domain knowledge through
  neuron-importance.
\newblock In {\em Proceedings of the European Conference on Computer Vision
  (ECCV)}, pp. 526--541, 2018.

\bibitem{selvaraju2017grad}
R.~R. Selvaraju, M.~Cogswell, A.~Das, R.~Vedantam, D.~Parikh, and D.~Batra.
\newblock Grad-cam: Visual explanations from deep networks via gradient-based
  localization.
\newblock In {\em Proceedings of the IEEE International Conference on Computer
  Vision}, pp. 618--626, 2017.

\bibitem{simonyan2013deep}
K.~Simonyan, A.~Vedaldi, and A.~Zisserman.
\newblock Deep inside convolutional networks: Visualising image classification
  models and saliency maps.
\newblock {\em ICLR}, 2014.

\bibitem{simonyan2014very}
K.~Simonyan and A.~Zisserman.
\newblock Very deep convolutional networks for large-scale image recognition.
\newblock {\em arXiv preprint arXiv:1409.1556}, 2014.

\bibitem{smilkov2017direct}
D.~Smilkov, S.~Carter, D.~Sculley, F.~B. Vi{\'e}gas, and M.~Wattenberg.
\newblock Direct-manipulation visualization of deep networks.
\newblock {\em ICML Workshop on Visualization for Deep Learning}, 2016.

\bibitem{smilkov2017smoothgrad}
D.~Smilkov, N.~Thorat, B.~Kim, F.~Vi{\'e}gas, and M.~Wattenberg.
\newblock Smoothgrad: Removing noise by adding noise.
\newblock {\em arXiv preprint arXiv:1706.03825}, 2017.

\bibitem{steiner2018impact}
D.~F. Steiner, R.~MacDonald, Y.~Liu, P.~Truszkowski, J.~D. Hipp, C.~Gammage,
  F.~Thng, L.~Peng, and M.~C. Stumpe.
\newblock Impact of deep learning assistance on the histopathologic review of
  lymph nodes for metastatic breast cancer.
\newblock {\em The American Journal of Surgical Pathology}, 42(12):1636--1646,
  2018.

\bibitem{sundararajan2017axiomatic}
M.~Sundararajan, A.~Taly, and Q.~Yan.
\newblock Axiomatic attribution for deep networks.
\newblock In {\em Proceedings of the 34th International Conference on Machine
  Learning-Volume 70}, pp. 3319--3328. JMLR. org, 2017.

\bibitem{szegedy2015going}
C.~Szegedy, W.~Liu, Y.~Jia, P.~Sermanet, S.~Reed, D.~Anguelov, D.~Erhan,
  V.~Vanhoucke, and A.~Rabinovich.
\newblock Going deeper with convolutions.
\newblock In {\em Proceedings of the IEEE Conference on Computer Vision and
  Pattern Recognition}, pp. 1--9, 2015.

\bibitem{van2009search}
F.~Van~Ham and A.~Perer.
\newblock “search, show context, expand on demand”: supporting large graph
  exploration with degree-of-interest.
\newblock {\em IEEE Transactions on Visualization and Computer Graphics},
  15(6):953--960, 2009.

\bibitem{wongsuphasawat2018visualizing}
K.~Wongsuphasawat, D.~Smilkov, J.~Wexler, J.~Wilson, D.~Mane, D.~Fritz,
  D.~Krishnan, F.~B. Vi{\'e}gas, and M.~Wattenberg.
\newblock Visualizing dataflow graphs of deep learning models in tensorflow.
\newblock {\em IEEE transactions on visualization and computer graphics},
  24(1):1--12, 2017.

\bibitem{zamir2018taskonomy}
A.~R. Zamir, A.~Sax, W.~Shen, L.~J. Guibas, J.~Malik, and S.~Savarese.
\newblock Taskonomy: Disentangling task transfer learning.
\newblock In {\em Proceedings of the IEEE Conference on Computer Vision and
  Pattern Recognition}, pp. 3712--3722, 2018.

\bibitem{zeiler2014visualizing}
M.~D. Zeiler and R.~Fergus.
\newblock Visualizing and understanding convolutional networks.
\newblock In {\em European Conference on Computer Vision}, pp. 818--833.
  Springer, 2014.

\end{thebibliography}
\end{document}